\documentclass[12pt,preprint]{aastex}

\newcommand\kms{km~s$^{-1}$}

\newcommand\etal{{et~al.}} 
\newcommand\mM{\ifmmode(m{-}M)\else$(m{-}M)$\fi}
\newcommand\msun{\ifmmode{\hbox{M$_\odot$}}\else{M$_\odot$}\fi}

\newcommand\hst{{\it HST}}

\newcommand\iz{\ifmmode(i_{775}{-}z_{850})\else$(i_{775}{-}z_{850})$\fi}
\newcommand\izacs{\ifmmode(i_{775}{-}z_{850})\else$(i_{775}{-}z_{850})$\fi}
\newcommand\ubrest{\ifmmode(U{-}B)_z\else$(U{-}B)_z$\fi}
\newcommand\bvrest{\ifmmode(B{-}V)_z\else$(B{-}V)_z$\fi}
\newcommand\zacs{\ifmmode z_{850}\else$z_{850}$\fi}
\newcommand\iacs{\ifmmode i_{775}\else$i_{775}$\fi}
\newcommand\vacs{\ifmmode V_{606}\else$V_{606}$\fi}
\newcommand\bacs{\ifmmode B_{435}\else$B_{435}$\fi}
\newcommand\zf{\ifmmode z_{\rm f}\else$z_{\rm f}$\fi}
\newcommand\lta{\lesssim}
\newcommand\gta{\gtrsim}
\newcommand\taul{\ifmmode\tau_{L}\else$\tau_{L}$\fi}

\shortauthors{Blakeslee et al.}
\shorttitle{Arc in Field Galaxy}
\tighten

\received{2003 October 28}
\begin{document}

\title{ACS Observations of a Strongly Lensed Arc in a Field Elliptical}

\author{John P.~Blakeslee\altaffilmark{1},
K.\ C.\ Zekser\altaffilmark{1},
N.\ Ben\'{\i}tez\altaffilmark{1},
M.\ Franx\altaffilmark{2},
R.\ L.\ White\altaffilmark{1,3},
H.\ C.\ Ford\altaffilmark{1},
R.~J.~Bouwens\altaffilmark{4},
L.\ Infante\altaffilmark{5},
N.\ J.\ Cross\altaffilmark{1},
G.\ Hertling\altaffilmark{5},
B.\ P.\ Holden\altaffilmark{4},
G.\ D.\ Illingworth\altaffilmark{4},
V.~Motta\altaffilmark{5},
F.\ Menanteau\altaffilmark{1},
G.\ R.\ Meurer\altaffilmark{1},
M.\ Postman\altaffilmark{1,3},
P.\ Rosati\altaffilmark{6},
W.~Zheng\altaffilmark{1}
}

\altaffiltext{1}{Department of Physics \& Astronomy, Johns Hopkins University, Baltimore, MD 21218; jpb@pha.jhu.edu}
\altaffiltext{2}{Leiden Observatory, P.O. Box 9513, 2300 Leiden, The Netherlands}
\altaffiltext{3}{Space Telescope Science Institute, 3700 San Martin Drive, Baltimore, MD 21218}
\altaffiltext{4}{Lick Observatory, University of California, Santa Cruz, CA 95064}
\altaffiltext{5}{Departmento de Astronom\'{\i}a y Astrof\'{\i}sica,
Pontificia Universidad Catolica de Chile, Santiago 22, Chile}
\altaffiltext{6}{European Southern Observatory, Karl-Schwarzschild-Str. 2, D-85748 Garching, Germany}

\begin{abstract}
We report the discovery of a strongly lensed arc system around a field elliptical
galaxy in {\it Hubble Space Telescope} (\hst) Advanced Camera for Surveys
(ACS) images of a parallel field observed during NICMOS observations of the
\hst\ Ultra-Deep Field.  The ACS parallel data comprise deep
imaging in the F435W, F606W, F775W, and F850LP bandpasses.
The main arc is at a radius of 1\farcs6 from the galaxy center
and subtends about 120$^\circ$.
Spectroscopic follow-up at Magellan Observatory yields a redshift
$z=0.6174$ for the lensing galaxy, and
we photometrically estimate $z_{\rm phot}=2.4\pm0.3$ for the arc.
We also identify a likely counter-arc at a radius
of 0\farcs6, which shows structure similar to that seen in the main arc.
We model this system and find a good fit to an elliptical
isothermal potential of velocity dispersion $\sigma \approx 300$ \kms,
the value expected from the fundamental plane, and some external
shear.  Several other galaxies in the field have colors similar
to the lensing galaxy and likely make up a small group.
\end{abstract}

\section{Introduction}

The past decade has witnessed a dramatic growth in both the number
of known gravitational lenses, and the astrophysical and cosmological
applications for which they have been used.
Strong lensing can provide valuable information about the 
detailed mass structure of intermediate-redshift galaxies 
(e.g., Rusin, Kochanek, \& Keeton 2003).
The frequency of lensing, as well as the
time delays and image positions as a function of redshift, 
can be used to constrain the 
cosmological parameters (e.g., Im \etal\ 1997, Chae \etal\ 2002).
Of course, lensing also provides
the opportunity to study the properties of distant objects
with the aid of strong magnification (e.g., Franx \etal\ 1997).

With  its unsurpassed resolution at optical wavelengths, 
the {\it Hubble Space Telescope} (HST), has played a central role in
the study and exploitation of many lensing systems.
The installation of the Advanced Camera for Surveys (ACS) 
(Ford \etal\ 2002), with its higher resolution, greater sensitivity,
and wider field, has made \hst\ a yet more powerful tool 
for gravitational lensing studies.
Here, we present our analysis of a strongly 
lensed  arc serendipitously discovered around a field
elliptical using ACS.  We adopt a cosmology with
$(h,\Omega_m,\Omega_\Lambda) = (0.7,0.3,0.7)$ throughout.

\vspace{-4pt}
\section{Observations and Data Reductions}

The object studied in this paper was found in a parallel field
observed with the ACS Wide Field Camera (WFC) during NICMOS observations of the 
\hst\ Ultra-Deep Field (GO program 9803, PI: Thompson) in 2003 September.  
It was observed  for 9,9,18,27 orbits in the F435W,F606W,
F775W,F850LP bandpasses, respectively, 
hereafter referred to as \bacs, \vacs, \iacs, and \zacs.
The pointing pattern was a grid of nine positions, spaced by about
one NICMOS field, or $\sim\,$30\arcsec.
These data are public and were retrieved from the STScI
archive where they received the standard CALACS on-the-fly
reprocessing (OTFR) to the point of flatfielded
images in units of electrons.  Further processing, including
cosmic ray rejection and final image combination, was performed with the
``Apsis'' data reduction software described in detail by Blakeslee \etal\
(2003a), with refinements as noted in Blakeslee \etal\ (2003b). 
Apsis now also performs automatic astrometric recalibration of the
images; the resulting zero-point error is $\sim\,$0\farcs1,
dominated by systematic effects in the astrometric catalogues.

Two of the \zacs\ exposures taken during one of the orbits were adversely
affected by a bright star that landed precisely at the edge of one of
the CCD chips, resulting in charge bleeding into the
overscan area used in the STScI CALACS processing.  This led to a large
erroneous slope in the fit to the overscan, making OTFR versions
of these images unusable.   We chose simply to omit these two images,
which reduced the final \zacs\ signal-to-noise by less than 2\%.
The final exposure times in our summed images were then
20700\,sec (\bacs),  20679\,sec (\vacs), 
41400\,sec (\iacs), and  59800\,sec (\zacs).
We use photometric AB magnitude zero points in these 
four bands of 25.673,  26.486,  25.654, and 24.862
(Sirianni \etal\ 2003, in preparation) and correct for 
$E(B{-}V)$ = 0.008 mag of extinction (Schlegel \etal\ 1998).
Finally, we obtained follow-up spectroscopy in 2003 November
with the Magellan Clay 6.5-m telescope using
the LDSS-2 at a dispersion of 5.3 \AA\,pix$^{-1}$.
The total exposure time was 7200 sec,
and the data were reduced using standard techniques.


\vspace{-4pt}
\section{Analysis and Results}

\begin{figure}\epsscale{0.97}
\plotone{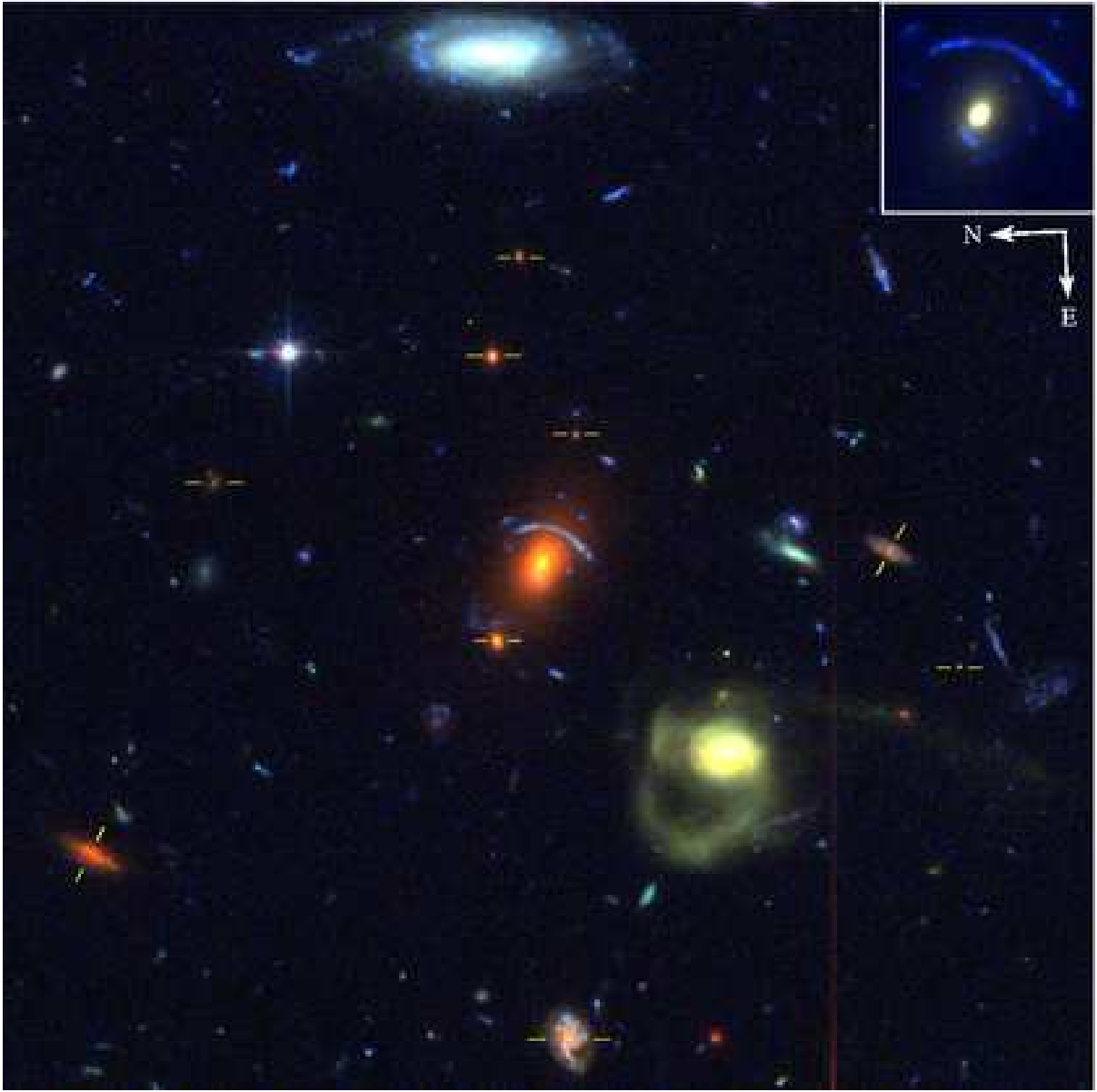}
\caption{ACS/WFC \bacs\vacs\iacs\ color composite image of a 0\farcm8 
square region centered on J033238$-$275653 and shown in the observed 
orientation. The prominent blue arc is $\sim\,$1\farcs6 W/SW of the lens galaxy; 
a candidate counter-arc and additional faint arc-like features 
(particularly to the immediate south of the elliptical)
are visible in the inset image.  Galaxies having photometric redshifts
with uncertainties $\lta0.1$ and values within $\pm0.1$ of the lens galaxy are marked.
The bright yellow-colored galaxy 12\arcsec\ to the SE of the lens
with extended tidal arms (reaching to the edge of the figure)
has a photometric redshift of 0.2.
\label{fig:colorim}}
\end{figure}

Figure~\ref{fig:colorim} shows a composite color image constructed
from a 0\farcm8 portion of the \bacs, \vacs, \zacs\ images.  A
dramatic blue arc, subtending $\sim\,$120$^\circ$ is readily visible
at a radius of 1\farcs63 perpendicular to the major axis of a red
early-type galaxy.  An apparent counter-arc of the same color at a
radius of 0\farcs6 can be seen in the inset image.  A number of other
blue, tangentially oriented features (e.g., a faint arc to the south
of the elliptical) may also be lensed.
Several nearby galaxies have colors and/or photometric
redshifts (see below) similar to the lensing galaxy  and likely
make up a group.  Keeton \etal\ (2000) predict that 20--25\% of galaxy
lenses will be in groups massive enough to perturb the lensing potential.
We note that this field has also been observed with Chandra (Rosati \etal\ 2002),
and we find an upper limit to the X-ray flux of 
$1{\,\times\,}10^{-15}$ erg cm$^{-2}\,$s$^{-1}$ in the 0.5--2 keV band, 
indicating $L_X < 2{\,\times\,}10^{42}$ erg\,s$^{-1}$ at 
the measured $z{\,=\,}0.62$.  The surrounding
group must then be  low-mass, $ \lesssim 2{\,\times\,}10^{13} M_\odot$,
assuming the $L_X{-}M$ relation from Reiprich \& B{\"o}hringer (2002),
although it could still have a significant effect on the lensing.

\begin{figure}\epsscale{0.54}
\plotone{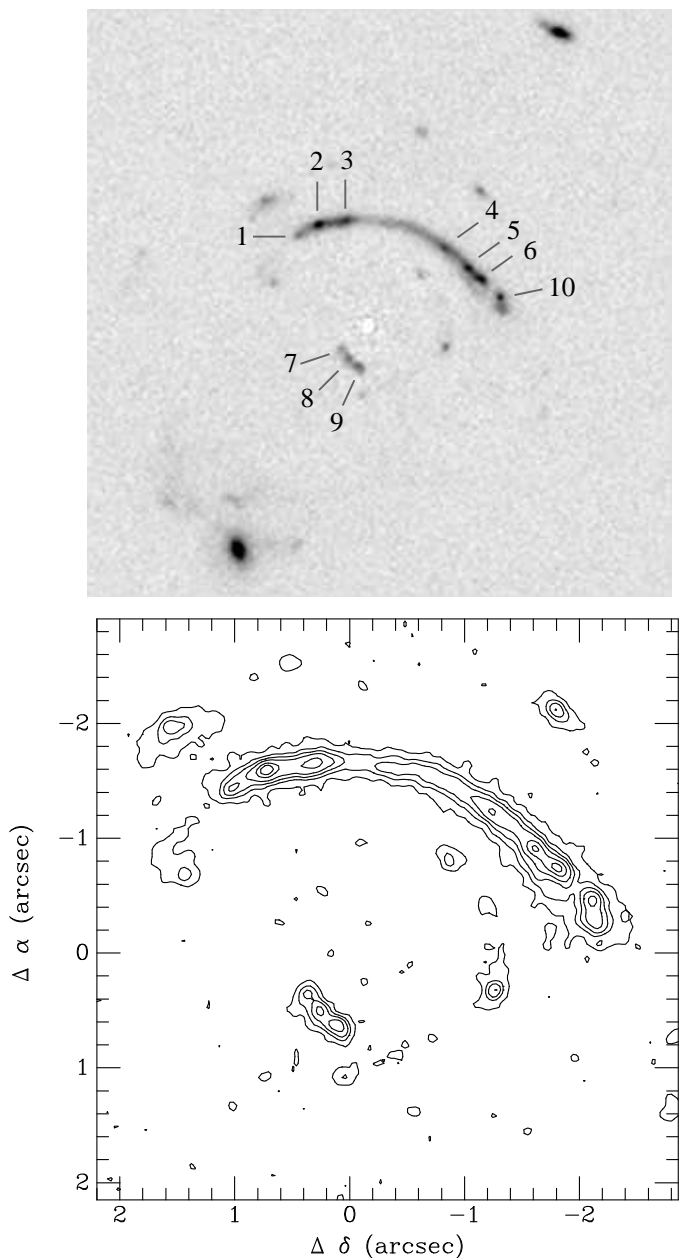}
\caption{A $9\arcsec{\times\,}9\arcsec$ region of the \bacs+\vacs\ image
(``drizzled'' to 0\farcs03\,pix$^{-1}$ to improve the sampling)
following galaxy model subtraction, with short lines labeling
several bright knots in the main arc and candidate counter-arc (top).
Contour plot of the central 5\arcsec\ of this region (bottom); most of the
objects have elongated morphologies or tangential orientations
suggestive of lensing.
%
For reference, the precise (RA,\,Dec) offsets 
from the galaxy center for knots 1--10  are, respectively:
$(-1\farcs46,  0\farcs97)$, $(-1\farcs62,  0\farcs65)$,
$(-1\farcs65,  0\farcs24)$, $(-1\farcs13, -1\farcs26)$,
$(-0\farcs79, -1\farcs60)$, $(-0\farcs61, -1\farcs78)$,
$( 0\farcs32,  0\farcs43)$, $( 0\farcs48,  0\farcs33)$,
$( 0\farcs61,  0\farcs20)$, $(-0\farcs30, -2\farcs06)$.
\label{fig:arcstruct}}
\end{figure}

For accurate photometric and structural measurements of the arc,
it is necessary to first model and subtract the light from the lensing
galaxy.  We used custom 2-d isophote-fitting software, masking the
arcs and other nearby sources before fitting.  The fitted models
resulted in small residuals and indicate an effective radius 
$R_e\approx1\farcs6$.
Figure~\ref{fig:arcstruct} illustrates the detailed knot structure that becomes
apparent in the arcs following galaxy subtraction;
the knot positions are given in the figure caption.
A combination of the \bacs\ and \vacs\ images is used
for this since the arcs are so blue relative to the galaxy.
The main arc is brighter than the counter-arc by a factor of $7.9\pm0.5$,
depending on whether or not the concentration near the south end of the
main arc is counted.  We tabulate our key measurements on this lensing
system in Table~\ref{tab:msmts}.  The galaxy redshift was measured from
the LDSS-2 spectra using template cross-correlation,
and the arc photometric redshift was estimated using ``BPZ'' 
(Ben\'{\i}tez 2000) with revised templates from Ben\'{\i}tez \etal\ (2004).

If we estimate the Einstein radius $\theta_E$ of the system to be
the projected distance of the primary arc
and assume a singular isothermal sphere model (e.g.,
Schneider, \etal\ 1992) with the redshifts from Table~\ref{tab:msmts},
then we derive a lens velocity dispersion $\sigma \approx 305$ \kms.  This
changes by only $\pm3\%$ if the source redshift is varied by $\mp0.4$. 
Interestingly, we predict a very similar $\sigma$ from a
fundamental plane (FP) analysis of the elliptical galaxy
using the data in Table~\ref{tab:msmts}.
We $K$-corrected our \zacs\ photometry to the SDSS $r$ band
and derived a value for $\sigma$ using the coefficients
and methods (including luminosity evolution of $-$0.85$\,z$ in
the $K$-corrected $r$ band and conversion of the fitted $R_e$ 
to an effective circular radius) from the very large field 
elliptical study by Bernardi \etal\ (2003).  The result is
$\sigma = 295\pm45$ \kms, where the error is based on the FP scatter.

We wish to investigate whether or not an elliptical galaxy of the
observed luminosity can naturally produce the observed lensed
features, including the counter-arc.
We construct a model using a round source galaxy and the singular
isothermal form of the elliptical effective lensing potential of
Blandford \& Kochanek (1987), parameterized by the velocity dispersion
$\sigma$ and an ellipticity $\epsilon_p$.
We also include an external shear component, which may arise from
a host group, the bright foreground galaxies seen in
Figure~\ref{fig:colorim}, or misalignment between the luminous
galaxy and dark matter halo (Keeton \etal\ 1997).

\begin{figure}\epsscale{0.9}
\plotone{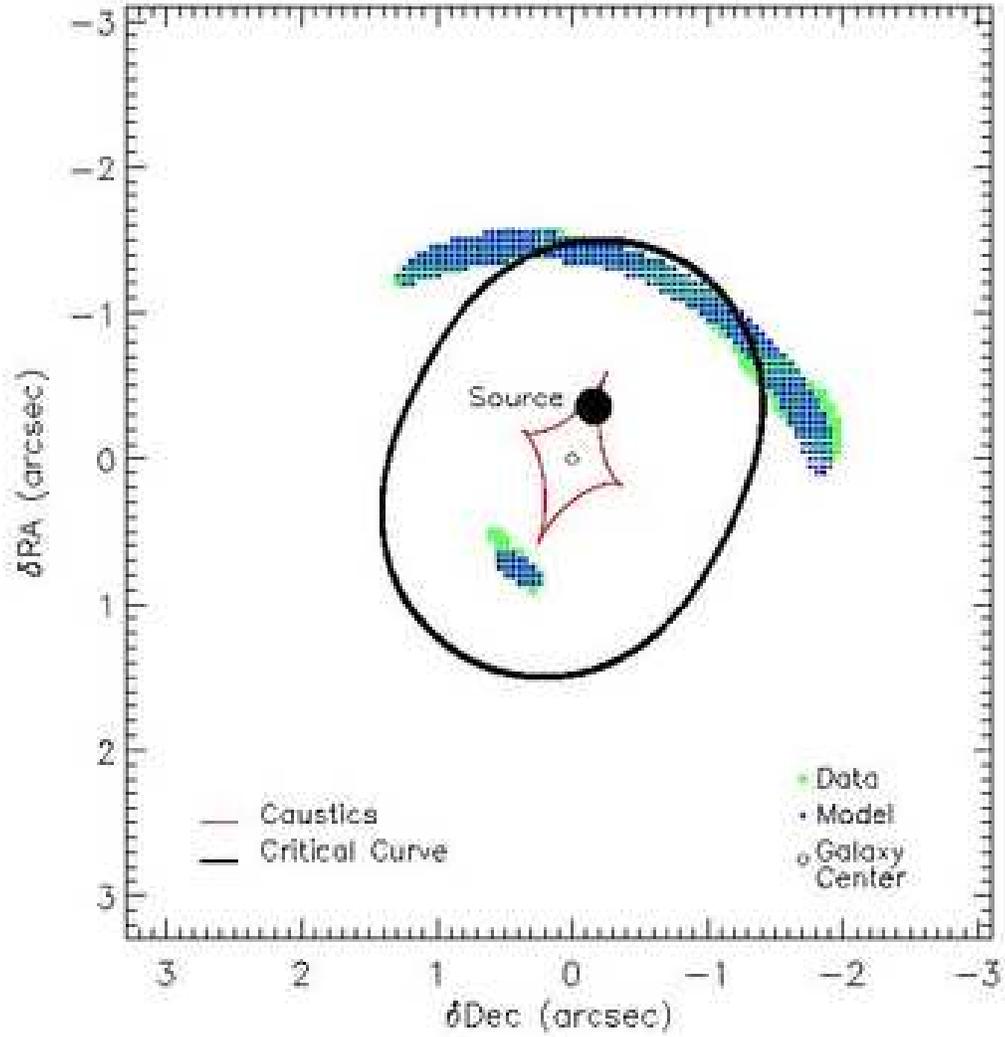}
\caption{Lensing model for the J033238$-$275653 system.
The galaxy center lies at (0,0), and the orientation is the same
as in Figures~\ref{fig:colorim} and \ref{fig:arcstruct}. 
The caustic and critical curves are
indicated in red and black, respectively.
The green shading illustrates the data
regions used in the modeling, while the small blue points indicate
the area to which the shown source is mapped in this model.
%
\label{fig:model}}
\end{figure}

Because the system is under-constrained, we fix some model parameters
to reasonable values.  In particular, we set $\sigma = 295$ \kms\ as
expected from the FP and $\epsilon_p = 0.12$, or a third of the
ellipticity of the light (Mellier \etal\ 1993).  We then allow the
position, orientation, and shear components to vary.
We used an optimization procedure that minimizes the deviation of the
delensed positions in the source plane while optimizing the gross
coincidence of the model and data in the image plane.  We did not
include magnification weighting or detailed constraints from the
internal structure of the arcs.  Figure~\ref{fig:model} shows that
the model successfully reproduces the observed geometry of the bright
arc and counter-arc using a single round source.  The major axis of
the mass ellipsoid is aligned to within 20\arcdeg\ of the 
galaxy light, while the external shear axis is within 10\arcdeg.
We defer more detailed modeling of the lens mass distribution and
internal structure of the arcs to a future work when we have
better information on the source redshift and lens velocity dispersion.

\vspace{-4pt}
\section{Discussion and Summary}

Early-type lensing galaxies constitute a small but
unique mass-selected sample in generally low-density
environments (Kochanek \etal\ 2000), lacking the biases inherent in
luminosity selection.  Kochanek \etal\ (2000) concluded that these
galaxies have old stellar populations with formation redshifts
$z_f\gta2$ and lie on the same FP as cluster ellipticals of the same
redshift. However, van de Ven \etal\ (2003) have analyzed the FP of 26
early-type lenses and find a range in colors and ages, with 
formation redshifts as low as $\sim\,$1 and slightly younger 
ages on average than cluster ellipticals.
The new lens studied here adds to the mass-selected
sample of field ellipticals, and the multi-band coverage 
allows for some constraint on age.  The observed 
$(\vacs{-}\iacs)$ and $(\iacs{-}\zacs)$ colors transform fairly
closely to rest-frame \ubrest\ and \bvrest, respectively.
We find $\ubrest = 0.45$, $\bvrest=0.85$, with $\sim \pm$0.03 mag
uncertainty in the transformations.  This implies an age 
$\tau_L = 4.4\pm0.7$ Gyr from the latest solar metallicity
Bruzual \& Charlot (2003) models, or formation at $z_f\approx2.0\pm0.5$.
Thus, the population is clearly evolved, though marginally younger
than derived for cluster ellipticals (e.g., van Dokkum \& Franx 2001;
Blakeslee \etal\ 2003b).

The excellent angular resolution and wide field of ACS on \hst\ 
now makes it possible to conduct systematic optical
searches for arcsecond-scale strongly lensed objects and 
obtain statistically meaningful results.  
The two Hubble Deep Fields observed with WFPC2
together produced only one probable lens candidate (Barkana \etal\ 1999),
while the medium-deep survey (MDS) found 10 good candidates 
over about 130 WFPC2 fields (Ratnatunga, Griffiths \& Ostrander 1999).
However, the recent paper by Fassnacht \etal\ (2003) lists 6 probable
lens candidates in ACS GOODS fields covering only about a quarter 
of the MDS area.  The dramatic instance of lensing we report here
lies in a field adjacent to the GOODS area and brings the number
of likely galaxy lenses found with ACS to 7, or about 40\% of the
total found with \hst\ in less than 15\% of its lifetime.
It seems likely that after a few more years, searches 
in ACS fields will provide the first
robust optical determination of the rate of strong gravitational 
lensing by individual galaxies, and thus their mass function.~~~ 

We have found and provided a first analysis of a clear case of strong
gravitational lensing by a field, or small group, elliptical galaxy.  
We estimate $z\approx2.4$ for the arc using our four-band ACS photometry
and have  spectroscopically measured $z=0.62$ for the lens galaxy.
Several smaller galaxies with similar colors 
may form a group with the lens galaxy,
although the Chandra X-ray flux limit indicates this must be fairly low-mass.
We have successfully modeled the arc/counter-arc system as a single round
source being lensed by an isothermal ellipsoid of velocity dispersion
$\sigma = 295$ \kms, the value indicated by the fundamental plane, with a
modest amount of external shear that could result from neighboring
or foreground galaxies.  
A more thorough analysis including detailed constraints on the mass
distribution and source morphology would benefit from measurements of the
arc redshift and lens galaxy velocity dispersion.
We expect these will be forthcoming, as will discoveries and analyses 
of many additional lensing systems with ACS.

\acknowledgments 

ACS was developed under NASA contract NAS 5-32864, and this research 
has been supported by NASA grant NAG5-7697.  
L.I., V.M., and G.H. are supported by 
a Chilean Conicyt Fondap grant ``Center for Astrophysics.''
We thank our fellow ACS Team members and support staff,
especially Dave Golimowski for retrieving the data
and Jon McCann for writing the software used in making Figure~1.

\clearpage

{{}
}

\begin{deluxetable}{lccc}\tablenum{1}
\centering
\tabletypesize{\small}
\tablecaption{J033238$-$275653 System Properties}\label{tab:msmts}
\tablecaption{Summary of Measurements}
\tablewidth{0pt} 
\tablehead{ 
\colhead{Quantity} & \colhead{value} & \colhead{$\;\pm$}& \colhead{units}
}\startdata
\multicolumn{4}{c}{Lens Galaxy} \\ \tableline
R.A.(J2000)  & 03:32:38.22           &$0\farcs1$ & h:m:s \\
Dec.(J2000)  & \llap{$-$}27:56:52.94 &$0\farcs1$ & \arcdeg:\arcmin:\arcsec  \\
total $z_{850,0}$  & 18.92 &  0.04 &  mag\\
$(\bacs{-}\vacs)_0$  & 2.48 & 0.02 &  mag \\
$(\vacs{-}\iacs)_0$  & 1.41 & 0.01 &  mag \\
$(\iacs{-}\zacs)_0$  & 0.43 & 0.01 &  mag \\
$R_e$  & 1\farcs60 & 0.25 & arcsec \\
ellipticity  & 0.36 & 0.05 & \dots \\
PA  & $58^\circ$ & $4^\circ$ & deg \\
$z_{\rm spec}$  & 0.6174 & 0.0003 & $\Delta \lambda/\lambda$ \\
\tableline
\multicolumn{4}{c}{Main Arc} \\ \tableline
total $V_{606,0}$  & 23.70 &  0.10 &  mag\\
$(\bacs{-}\vacs)_0$  & 0.33 & 0.02 &  mag \\
$(\vacs{-}\iacs)_0$  & 0.10 & 0.01 &  mag \\
$(\iacs{-}\zacs)_0$  & \llap{$-$}0.01 & 0.02 &  mag \\
radius  & 1\farcs63 & 0.02 & arcsec \\
length  & \llap{$\gta\,$}3\farcs35 & \dots & arcsec \\
$z_{\rm phot}$  & 2.4 & 0.3 & $\Delta \lambda/\lambda$ \\
\enddata
\vspace{-16pt}
\tablecomments{All magnitudes are AB; galaxy colors
are within a 0\farcs5 radius; PA is measured E from N.
Arc ``radius'' refers to separation along major axis.  
}
\end{deluxetable}

\end{document}